\documentclass[aps,prf,print]{revtex4-2}
\usepackage{fancyhdr}
\usepackage{comment}
\usepackage{graphicx}
\usepackage{newtxtext}
\usepackage{newtxmath}
\usepackage{natbib}
\usepackage{hyperref}
\usepackage{nomencl}
\usepackage{ifthen}
\renewcommand{\nomgroup}[1]{%
\ifthenelse{\equal{#1}{C}}{\item[\textbf{Constants}]}{%
\ifthenelse{\equal{#1}{O}}{\item[\textbf{Overscript}]}{%
\ifthenelse{\equal{#1}{S}}{\item[\textbf{Subscript}]}{}}}
}
\hypersetup{
   colorlinks = true,
    urlcolor   = blue,
    citecolor  = black,
    linkcolor = blue,
}

\newcommand{\RomanNumeralCaps}[1]
\linenumbers
\usepackage[noabbrev]{cleveref}
\begin{document}

\title{On the early stages of vapor bubble growth: From the surface-tension to the inertial regime}
\author{Orr Avni}
\affiliation{Faculty of Aerospace Engineering, Technion - Israel Institute of Technology, Haifa, 320003, Israel.}
\author{Eran Sher}
\affiliation{Faculty of Aerospace Engineering, Technion - Israel Institute of Technology, Haifa, 320003, Israel.}
\author{Yuval Dagan}
\email{yuvalda@technion.ac.il}
\affiliation{Faculty of Aerospace Engineering, Technion - Israel Institute of Technology, Haifa, 320003, Israel.}
\date{\today}
\begin{abstract}
This paper presents a new analytical model for the early stages of vapor bubble growth in superheated liquids.
The model bridges a gap in current knowledge by focusing on the surface tension-controlled, near-equilibrium growth regime and its transition to an inertia-controlled regime.
A unified analytical model is derived by combining a perturbation method for the initial growth regime with a complementary outer solution to model the subsequent bubble growth rate. 
The model successfully predicts the initial delay in bubble growth due to surface tension effects.
Two non-dimensional parameters govern this delay period: the initial perturbation from the equilibrium radius and the Ohnesorge number at the onset of nucleation.
The Ohnesorge number encapsulates the interplay between viscous damping and surface tension forces acting on the bubble during its early growth stages. 
The analytical solutions presented here allow for quantifying the surface-tension, inertial, and transitional regimes while establishing a simple criterion for estimating the influence of thermal effects on early-stage growth.
Our findings emphasize the significance of considering the surface tension delay, particularly for short timescales. 
The derived analytical solutions and the obtained correlation for surface-tension-induced delay may prove a practical tool and could be integrated into existing models of vapor bubble growth.
\end{abstract}
\maketitle
\thispagestyle{fancy}
\section{Introduction}\label{sec:intro}
Nucleation of vapor bubbles within a superheated liquid is a fundamental yet complex and elusive phase-change phenomenon.
The ubiquity and role vapor bubbles play in natural processes \citep{Gardner2012,Vincent2012}, as well as wide-ranging industrial \citep{Apfel1979,Avedisian1999,DiFulvio2013,Avni2022,BarKohany2023a}, and medical \citep{Maxwell2011,Bader2019,Mancia2020} applications poses a solid scientific research motivation.
Embryonic vapor nucleons constantly form spontaneously in superheated liquids due to the liquid's inherent local density fluctuations; the contrasting effects of the inward-directed surface tension and outward-directed pressure acting on the nucleated bubble interface cause small nucleons to collapse immediately \citep{Debenedetti1996}.
However, nucleons surpassing the unstable equilibrium radius initiate localized liquid-vapor phase change.
The subsequent growth of a single vapor in an unbounded superheated liquid includes three discernible stages \citep{Plesset1977,Prosperetti2017}.
Surface tension forces predominate the initial growth stage, impeding significant growth for a -- usually short -- delay period \citep{Robinson2004}.
Following the growth of the nucleon, surface tension influence diminishes, and inertia forces dominate the growth.
The constant growth rate is then driven by a steady pressure gradient across the interface, originating from the liquid's superheat degree \citep{Rayleigh1917}.
The growth transitions to the last regime as heat consumed by evaporation at the interface cools the liquid boundary layer around it, reducing the vapor pressure inside the bubble.
As the pressure gradient tends to zero, the growth is driven solely by heat supplied to the interface via thermal diffusion; hence, during this asymptotic stage, the growth rate is proportional to the square root of time \citep{Plesset1954}. 

Most current analytical models focus on the latter stages of the growth; notably, the Mikic–Rohsenow–Griffith (MRG) model \citep{Mikic1970} pioneered the interpolation between the inertial and thermal regimes.
Following works \citep{Board1971,Theofanous1976,Prosperetti1978} have since modified the interpolation and improved its accuracy in the transitional regime. 
The surface-tension regime was addressed by \citeauthor{miyatake1997}~\citep{miyatake1997}, proposing a simple model considering bubble growth acceleration effects, albeit with a delay fitted using experimental data without assessing the influence of viscous damping effects.
Recently, \citeauthor{Sullivan2022}~\citep{Sullivan2022} presented an improved interpolated semi-analytical model capable of approximating the growth of the bubble across all three regimes.
The study involved a numerical investigation of near-equilibrium growth, subsequently integrated into a broader analytical framework for comprehensively modeling bubble growth.

\textcolor{black}{On the other hand, numerical studies \citep{Lee1996,Hao1999,Robinson2004}} solved the coupled inertial-thermal problem and successfully captured the entirety of the growth process, including the surface tension regime.
\textcolor{black}{Another approach to investigate the growth of a post-nucleation vapor bubble is molecular dynamics simulations.}
Using Lennard-Jones large computational volume simulations of liquid-to-vapor nucleation, \citeauthor{Angelil2014}~\citep{Angelil2014} managed to simulate the properties of bubbles from their inception as stable, critically sized bubbles to their continued growth.
Crucially, the study reported good agreement up to the molecular scale between molecular dynamics simulations and results obtained via the simplified Rayleigh-Plesset model used in most analytical and numerical models.
However, both numerical and molecular-dynamics-based works are limited in their generality and applicability compared with analytical models, which yield immediate predictions without requiring significant computational effort.
Studying the early dynamics of vapor bubbles faces experimental challenges; the vapor bubble critical radius may reach extremely small scales ~\citep{Avni2023}, especially when considering nucleation in highly superheated liquids. 
Under such conditions, capturing substantial stages of the bubble's growth -- especially the initial surface-tension-dominated period -- could be limited by current photography techniques.
Therefore, this study focuses on analytical modeling of the near-equilibrium, surface-tension-dominated growth regime and its transition to an inertia-dominated regime, aiming to bridge the existing theory gap.
We adopt a regular perturbation method to derive an approximate near-equilibrium solution; the inner solution is then matched with a solution for the transitional regime, yielding a unified model. 
Using this method, one may explore the underlying physics governing the dynamics of vapor bubbles following nucleation inception, aiming to provide a better understanding of their early growth stages.

\Cref{sec:model} outlines the methodology and mathematical modeling approach used to derive an analytical solution for near-equilibrium bubble growth and a complementary outer solution for the transitional regime from surface-tension-controlled to inertia-controlled growth. Selected results of the model are presented in \cref{sec:results}, followed by an outlook of the present analysis in \cref{sec:conc}.

\section{Mathematical Model}\label{sec:model}
The immediate aftermath of a spontaneous nucleation event is considered here: the growth of a vapor bubble driven by a phase change within a superheated liquid.
Such vapor-liquid systems are mechanically unstable; the equilibrium bubble radius is predicted by the Young-Laplace relation  
\begin{equation}
    R_c = \frac{2\sigma}{\Delta p_0},
\end{equation}
where $R_c$ is the critical bubble radius, $\sigma$ is the vapor-liquid surface tension, and $\Delta p_0$ is the pressure gradient acting on the interface.
Thus, we consider an initially still bubble embryo whose radius is perturbed with respect to the critical radius, $\tilde{R} \left({\tilde{t}}=0\right) = R_c + \tilde{\varepsilon} $.
Since the equilibrium embryo distribution $n_e (R)$ is related exponentially to the minimal work required for its formation \citep{Debenedetti1996,Blander1975},
\begin{equation}
    n_e (\tilde{R}) \propto \exp \left[ {-\frac{4 \pi \sigma R_c^2}{3kT} \left(1 -  \frac{3(\tilde{R}-R_c)^2}{R_c^2} \right) }\right],
\end{equation}
this perturbation is associated with the thermodynamic density fluctuations that lead to the bubble's nucleation.
Far from the critical point, the naturally occurring perturbations are in the molecular order of magnitude.

Thus, in this study, we postulate that the typical size of the physical perturbations is in the range of $\tilde{\varepsilon} = 0.1~nm$ to $\tilde{\varepsilon} = 10~nm$. 
Furthermore, the model assumes a perfectly spherical bubble immersed in a semi-infinite, incompressible liquid medium; thus, any potential interactions between nucleated bubbles are negligible.
Our analysis is restricted to early growth stages, wherein thermal diffusion effects are insignificant and constant temperature difference across the sharp interface is maintained.
The extent to which this approximation is valid will be further discussed in \cref{sec:unif}.
Finally, we assume the bubble does not contain dissolved gases and consists solely of vapor molecules, maintaining thermodynamic equilibrium with the outer superheated liquid. 
Given a constant vapor temperature, a saturated state implies that the growth process is driven by a steady pressure gradient across the bubble interface $\Delta p_0$.
The steady driving force distinguishes the growth process of a vapor bubble from the dynamics of a gas bubble, wherein the inner pressure is coupled to the bubble's volume.

The simplifying assumptions yield a single ordinary differential equation for the bubble radius $\tilde{R}$,  the Rayleigh-Plesset (RP) equation \citep{Plesset1954};
\begin{equation}\label{eq:RP}
    \tilde{R} \ddot{\tilde{R}} + \frac{3}{2}{\dot{\tilde{R}}}^2 = \frac{\Delta p_0}{\rho_l} -\frac{2}{\rho_l \tilde{R}}\left ( {\sigma +2\mu_l \dot{\tilde{R}}} \right) ,
\end{equation}
where $\rho_l$ is the liquid's density and $\mu_l$ is its viscosity.
The resulting bubble oscillator experiences external forcing from a constant pressure gradient, countered by surface tension spring-like rigidity and the damping effect of viscosity.
The inertia terms of the oscillator are nonlinear, originating from the integration of the unsteady (first LHS term) and radially convective (second LHS term) components of the Navier-Stokes equation. 
Using the critical radius $R_c$ and the inertial regime velocity
\begin{equation}\label{eq:RdotINER}
    {\dot{\tilde{R}}}_{IN} = \sqrt{\frac{2\Delta p_0}{3\rho_l}} = \sqrt{\frac{4\sigma}{3\rho_l R_c}},
\end{equation}
we normalize \cref{eq:RP} and its initial conditions, yielding the following initial value problem (IVP)
\textcolor{black}{\begin{equation}\label{eq:RPnd}
   \frac{2}{3} R \ddot{R} + {\dot{R}}^2 -1  + \frac{1}{R}\left(1+ \frac{4}{\sqrt{3}}\text{Oh}_{c}\dot{R}\right)  = 0; \hspace{0.5cm} {R}(0) = 1+\varepsilon; \hspace{0.5cm} \dot{R}(0) = 0,
\end{equation}
where $\text{Oh}_{c}$ is the \textit{nucleation Ohnesorge number}
, defined as
\begin{equation}\label{eq:Oh}
    \text{Oh}_{c} = \frac{\mu_l}{\sigma}\sqrt{\frac{\Delta p_0 }{ 2\rho_l }} =\sqrt{\frac{\mu_l^2}{ \sigma \rho_l R_c}},
\end{equation}}
and the normalized initial perturbation is
\begin{equation}
    \varepsilon = \frac{\tilde{\varepsilon}}{R_c}.
\end{equation}
The nucleation Ohnesorge number \textcolor{black}{$\text{Oh}_{c}$} signifies the ratio between viscous damping and surface tension forces acting on the critically sized bubbles, i.e., the balance between the two at -- and shortly after -- nucleation inception.
Since the surface tension forces are coupled to the bubble's characteristic radius, the nucleation Ohnesorge number is governed by both liquid medium properties and the thermodynamic conditions at nucleation.

\subsection{Near-equilibrium growth}\label{sec:inner}
The inherent nonlinearity of the RP equation restricts the attainment of a fully analytical solution to reduced, limiting cases \citep{bender1999a}.
Hence, we first seek an approximate solution for the initial, near-equilibrium growth of the vapor bubble by substituting a regular perturbation series of the bubble radius 
\begin{equation}
    R = R_0 + \varepsilon R_1 + \varepsilon^2 R_2 + \varepsilon^3 R_3 +...
\end{equation}
in the original IVP. 
The leading order solution is a trivial one $R_0 = 1$, indicating the stability of the bubble in the absence of any perturbation.
The first-order approximation yields a linearized, homogeneous IVP
\begin{equation}\label{eq:ode1}
   \mathcal{L}\left[R_1\right]=\frac{2}{3}{\ddot{R}}_1 + \frac{4}{\sqrt{3}}\textcolor{black}{\text{Oh}_{c}}{\dot{R}}_1 - R_1 = 0; \hspace{0.5cm} {R}_1(0) = 1;\hspace{0.5cm}\dot{R}_1(0) = 0,
\end{equation}
and the subsequent terms in the perturbation series yield the following heterogeneous IVP:
\begin{equation}\label{eq:ode2}
    \mathcal{L}\left[R_2\right]= -\frac{4}{3}R_1\ddot{R_1} - \dot{R_1}^2; \hspace{0.5cm}{R}_2(0) = \dot{R}_2(0) = 0, \\
\end{equation}
and
\begin{equation}\label{eq:ode3}
    \mathcal{L}\left[R_3\right]= -\frac{4}{3}\left(R_1\ddot{R_2}+R_2\ddot{R_1}\right)-2\dot{R_1}\dot{R_2} - R_1\dot{R_1}^2 - \frac{2}{3}R_1^2\ddot{R_1}; \hspace{0.5cm}{R}_3(0) = \dot{R}_3(0) = 0.
\end{equation}

The particular solution for the first-order approximation is
\begin{equation}\label{eq:sol1}
   R_1 (t) = e^{-\sqrt{3} t \text{Oh}_c } \left[\cosh\left(\sqrt{3} t \sqrt{\frac{1}{2} +  \text{Oh}_c^2}\right) + \frac{\text{Oh}_c}{\sqrt{\frac{1}{2} + \text{Oh}_c^2} } \sinh\left(\sqrt{3} t \sqrt{\frac{1}{2} +  \text{Oh}_c^2}\right) \right],
\end{equation}
which may be substituted into \cref{eq:ode2},
\begin{equation}
    \frac{2}{3}{\ddot{R}}_2 + \frac{4}{\sqrt{3}}\textcolor{black}{\text{Oh}_{c}}{\dot{R}}_2 - R_2= \frac{e^{-2\sqrt{3} t \text{Oh}_c}}{4+  8\text{Oh}_c^2} \left[1+ 16  \textcolor{black}{\text{Oh}_c^2}  + 7 \cosh\left(2\sqrt{3} t   \sqrt{\frac{1}{2} +  \text{Oh}_c^2}\right)\right], \\
\end{equation}
to yield the second-order approximation,
\begin{equation}
    R_2(t) = \frac{\sqrt{3}}{24} e^{-3\sqrt{3} t \text{Oh}_c} \sinh\left(\sqrt{3} t \sqrt{\frac{1}{2} + \text{Oh}_c^2}\right) \left[1 + 16 \text{Oh}_c^2 + 7 \cosh\left(2\sqrt{3} t \sqrt{\frac{1}{2} + \text{Oh}_c^2}\right)\right].
\end{equation}
The linearity of the problem allows for the derivation of higher-order approximations using the same methodology.
The perturbation solution is, by definition, limited to near-equilibrium bubble radii.
However, in the subsequent section, we derive a complementary analytical solution using different approximations to model the growth of the bubble during its transition from the surface-tension-dominated, near-equilibrium regime to the inertial regime.

\subsection{Transition to inertial growth} \label{sec:transition}
\begin{figure*}
    \centering 
    \includegraphics[width=\linewidth]{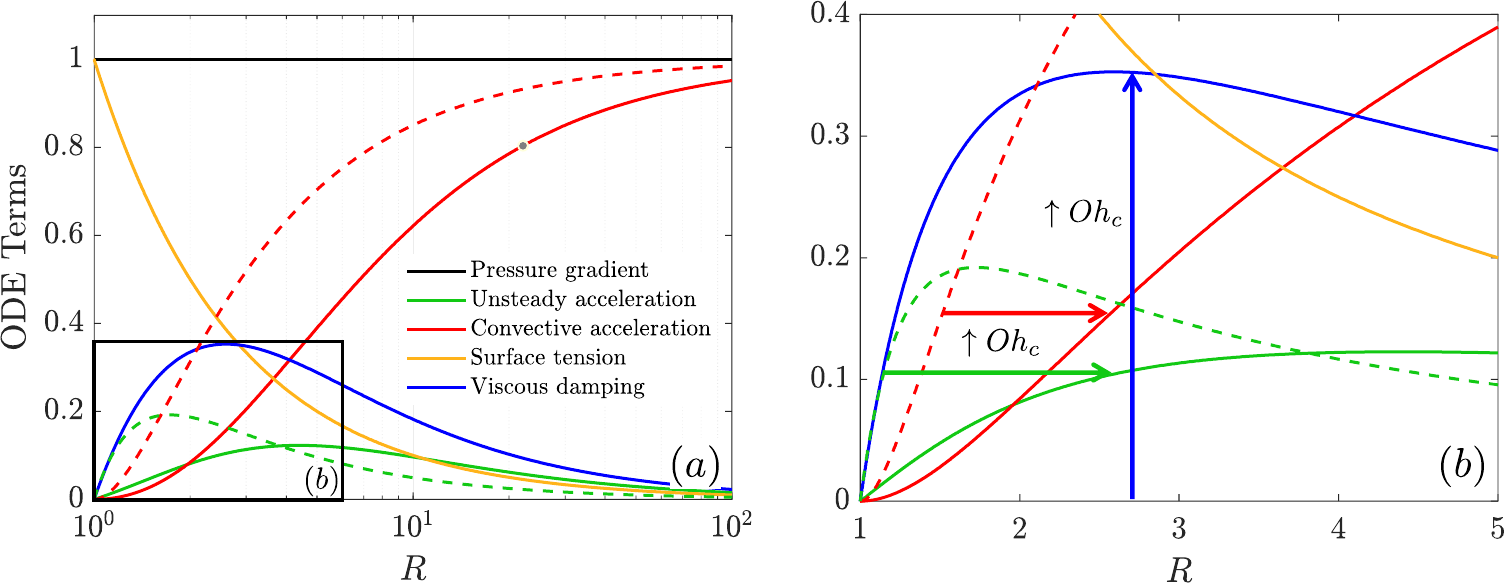}
    \caption{Dominant balance analysis of the normalized RP equation using a numerical solution for \cref{eq:RPnd}, given initial radius perturbation of $\varepsilon = 10^{-5}$. 
(a) Solid lines depict a test case \textcolor{black}{$\text{Oh}_{c} = 1$}, whereas the dashed lines depict the results for undamped growth \textcolor{black}{$\text{Oh}_{c} =0$}.  
(b) Close-up of the initial growth stages. Changes due to increased damping are highlighted with colored arrows.}
    \label{fig:DB}
\end{figure*}

\Cref{fig:DB} presents a dominant balance analysis of the nonlinear oscillator leading terms based on the numerical solution of \cref{eq:RPnd} using an RK4 scheme.
Each term of the normalized RP equation is plotted individually, allowing one to assess the dominant physical terms throughout the growth. 
The analysis is conducted for two distinct test cases: generalized, viscosity-affected growth wherein $\text{Oh}_{c} = 1$ (solid lines) and a limit case of undamped growth, wherein $\text{Oh}_{c} \rightarrow 0$ (dotted lines).
As our model postulates, the normalized pressure term remains constant in both cases; the surface tension and convective inertia terms switch roles as the second dominant terms.
The linearized \cref{eq:ode1} and \cref{fig:DB} suggest the initial growth is governed by the balance between pressure and surface tension forces, justifying the omission of the nonlinear convective inertia term in the linearized IVP.
However, as the bubble interface accelerates, the nonlinear term becomes dominant, breaking the assumptions of our linear analysis and inducing a shift in the bubble growth regime.
A distinct velocity threshold emerges in both cases, delineating the crossing from a surface-tension-dominated to a transitional regime, where both inertia and surface-tension forces govern the growth.
Viscous effects also influence the dominant terms; they delay the surge in the inertia term, playing a significant role immediately post-nucleation and up until the bubble grows beyond the critical radius order of magnitude $(R-1)>1$.
Viscous damping also affects the transitional regime; however, its magnitude diminishes as the bubble radius increases.
Thus, it does not influence the inertial regime, wherein the dominant terms are the pressure gradient and the convective acceleration.

The dominant balance analysis also indicates that the unsteady inertia term influence is comparatively insignificant in the system response.
This observation allows for extracting a reduced outer solution for the intermediate and inertial regime by excluding the unsteady inertia term. \Cref{eq:RPnd} simplifies into an autonomous, first-order ODE,
\begin{equation}
    R\left({{\dot{R}}^2-1}\right) + 1 + \frac{4}{\sqrt{3}}\text{Oh}_{c}\dot{R} = 0,
\end{equation}
amenable to a solution by inversing the ODE
\begin{equation}\label{eq:outersolODE}
    {\left(\frac{dt}{dR}\right)}^2 - \frac{4\text{Oh}_{c}}{\sqrt{3}(R-1)}\frac{dt}{dR} - \frac{R}{R-1}=0.
\end{equation}
One may solve for $t(R)$ given $R>1$, yielding 
\begin{multline}\label{eq:outer}
    t(R) = C_1 + \sqrt{R(R-1) + \frac{4\text{Oh}_c^2}{3}} + \frac{4\text{Oh}_c}{\sqrt{3}}\ln{\left(2 - 2R - \frac{4\text{Oh}_c}{\sqrt{3}} + 2 \sqrt{R(R-1) + \frac{4\text{Oh}_c^2}{3}} \right)} \\
    - \frac{1}{2}\left(1+\frac{4\text{Oh}_c}{\sqrt{3}}\right) \ln{\left(1 - 2R + 2\sqrt{R(R-1) + \frac{4\text{Oh}_c^2}{3}}\right)}
\end{multline}
However, the remaining degree of freedom $C_1$ cannot be independently resolved due to the lack of boundary condition.
Nevertheless, we may extract a uniform solution by imposing continuity of both radius and velocity terms with the inner solution derived in \cref{sec:inner}; for simplicity, we use here the first-order solution \cref{eq:sol1}.
Velocity continuity necessitates that the solutions are tangent $\dot{R}_{in}=\dot{R}_{out}$ at their intersection, resulting in a closed-form algebraic system for the tangent intersection point $\left(t_{tan},R_{tan}\right)$ and the integration constant $C_1$ as a function of $\text{Oh}_c$ and the $\varepsilon$
\begin{multline}\label{eq:Rtan}
    \frac{1}{2}\sqrt{\frac{3}{2}+3\text{Oh}_{c}^2} \varepsilon e^{-\sqrt{3}t_{tan} \text{Oh}_{c} } \sinh\left(\sqrt{3} t_{tan} \sqrt{\frac{1}{2} + \text{Oh}_{c}^2}\right) = \left({\frac{2 \text{Oh}_{c}}{\sqrt{3} (R_{tan}-1)} + \sqrt{\frac{2 \text{Oh}_{c}^2 - 3R_{tan}}{3R_{tan} - 3}}}\right)^{-1},\\
    R_{tan} = 1 + \varepsilon e^{-\sqrt{3} t_{tan} \text{Oh}_c } \left[\cosh\left(\sqrt{3} t_{tan} \sqrt{\frac{1}{2} +  \text{Oh}_c^2}\right) + \frac{\text{Oh}_c}{\sqrt{\frac{1}{2} + \text{Oh}_c^2} } \sinh\left(\sqrt{3} t_{tan} \sqrt{\frac{1}{2} +  \text{Oh}_c^2}\right) \right],\\
    C_1 = t_{tan}  - \sqrt{R_{tan}(R_{tan}-1) + \frac{4\text{Oh}_c^2}{3}} - \frac{4\text{Oh}_c}{\sqrt{3}}\ln{\left(2 - 2R_{tan} - \frac{4\text{Oh}_c}{\sqrt{3}} + 2 \sqrt{R_{tan}(R_{tan}-1) + \frac{4\text{Oh}_c^2}{3}} \right)} + \\ \frac{1}{2}\left(1+\frac{4\text{Oh}_c}{\sqrt{3}}\right) \ln{\left(1 - 2R_{tan} + 2\sqrt{R_{tan}(R_{tan}-1) + \frac{4\text{Oh}_c^2}{3}}\right)},
\end{multline}
which fully resolves the outer solution by setting $C_1$.

\section{Results and discussion}\label{sec:results}

We begin by analyzing the inner, near-equilibrium solution, focusing on the immediate aftermath of nucleation events and the early stages of bubble growth. 
The subsequent section presents results for the unified model, broadening the model's scope and shedding light on the transition of the growth to the initial regime.

\subsection{Surface tension delay time}\label{sec:tau_ST}
The accuracy and validity region of the approximated solutions is assessed by comparing them with numerical solutions obtained from the original RP oscillator equation \cref{eq:RPnd}.
An exemplary test case for generalized damped growth is considered, where the conditions at nucleation yield an Ohnesorge number of $\text{Oh}_c=1$ and an initial perturbation of $\varepsilon = 10^{-5}$.
\begin{figure*}
    \centering 
    \includegraphics[width=\linewidth]{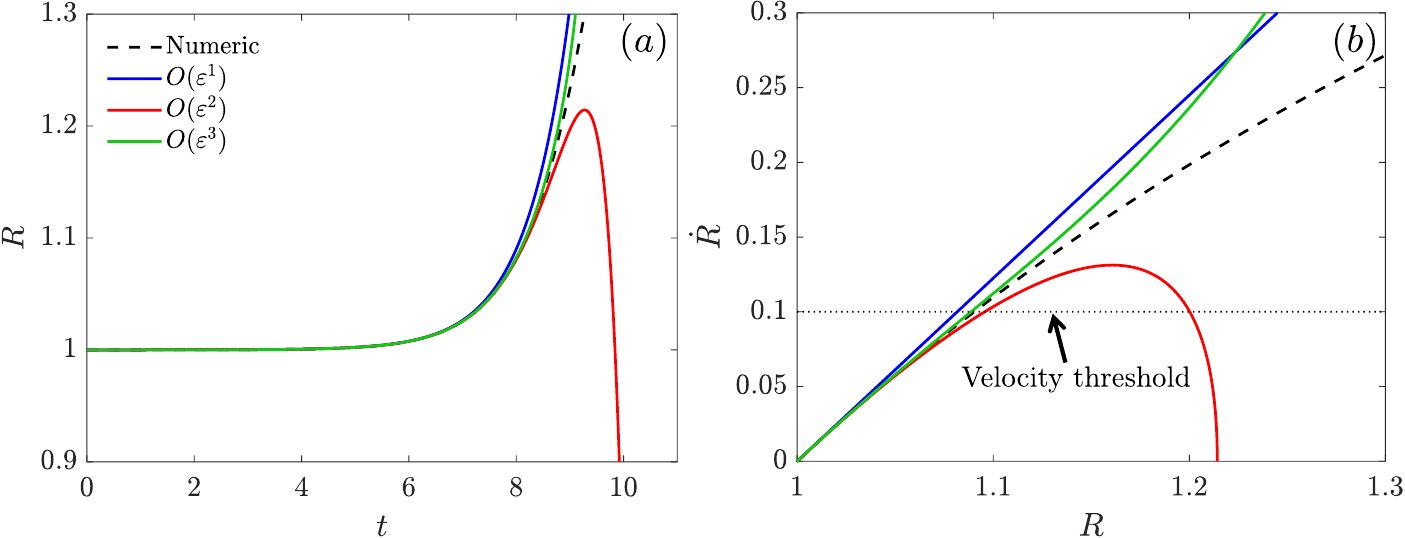}
    \caption{A comparison between the numerical solution of the RP equation (dashed line) and different orders of the approximated perturbation solution (solid lines), given an initial radius perturbation of $\varepsilon = 10^{-5}$ and $\text{Oh}_c=1$. (a) Bubble radius as a function of time and (b) Bubble interface velocity as a function of bubble radius}
    \label{fig:UnordCOMP}
\end{figure*}
\Cref{fig:UnordCOMP} compares successive orders of the inner solution with a numerical solution. 
The approximated and numerical solutions converge and are indiscernible at the initial growth stages; as the bubble grows, the approximated solutions diverge as the hyperbolic terms in each approximation asymptotically approach $\pm \infty$.
A higher order of approximation leads to improved precision and a marginally expanded region of validity.
Nevertheless, \cref{fig:UnordCOMP}(b) reveals a divergence among all solutions emerges when the velocity attains a magnitude of $\dot{R}\approx0.1$, i.e., when the velocity reaches the inertial velocity order of magnitude.
This divergence across all solutions accentuates the critical influence of the nonlinear convective inertia term $\dot{R}^2$ on the near-equilibrium bubble dynamics and the transition between the distinctly different growth regimes, as suggested by \cref{fig:DB}.

\begin{figure}
    \centering 
    \includegraphics[width=.6\columnwidth]{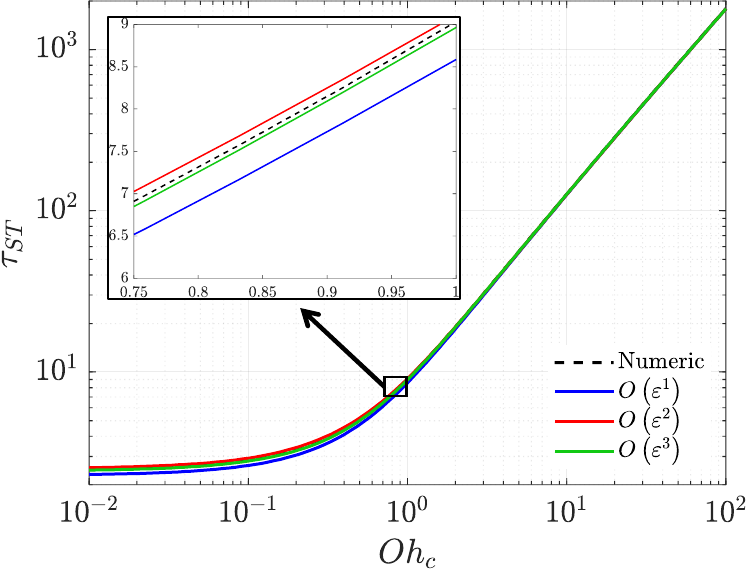}
    \caption{Surface-tension-induced delay as a function of the nucleation Ohnesorge number \textcolor{black}{$\text{Oh}_{c}$}, given initial radius perturbation of $\varepsilon = 10^{-5}$. Comparison between results obtained using the numerical solution, represented by the dashed black line, and results obtained using successive orders of the approximated perturbation solution, represented by solid lines.}
    \label{fig:DampedordCOMP}
\end{figure}

The surface tension delay, i.e., the time between nucleation onset and the crossing of the velocity threshold, can be evaluated using the derived inner solution. \Cref{fig:DampedordCOMP} presents the surface-tension-induced delay as a function of the nucleation Ohnesorge number \textcolor{black}{$\text{Oh}_{c}$} for successive orders of the perturbation solution, comparing them with the prediction of the numerical solution.
We observe varying sensitivity to the approximation order; low nucleation Ohnesorge values exhibit greater sensitivity to the approximation order, whereas the results converge for \textcolor{black}{$\text{Oh}_{c}>1$}.
A distinct transition of the initial delay behavior as a function of nucleation Ohnesorge number is illuminated by \cref{fig:DampedordCOMP}. 
For small values of nucleation Ohnesorge number, the surface tension delay remains unaffected by changes in its value. 
As the nucleation Ohnesorge number tends to \textcolor{black}{$\text{Oh}_{c} =0.1$}, the delay begins to increase, marking a shift to a linear relation between the two for \textcolor{black}{$\text{Oh}_{c} > 1$}.
This transition indicates a change in the dominant delaying mechanism: for \textcolor{black}{$\text{Oh}_{c}<0.1$}, the initial delay is primarily governed by surface tension, rendering it independent of the nucleation Ohnesorge number.
Contrarily, for \textcolor{black}{$\text{Oh}_{c}>1$}, viscous damping becomes the dominant factor setting the initial delay, manifested by the obtained linear correlation between \textcolor{black}{$\text{Oh}_{c}$} and $\tau_{ST}$.

Although the least accurate, \cref{fig:UnordCOMP} and \cref{fig:DampedordCOMP} suggest that the first-order solution captures the essence of the initial delay while yielding a closed-form relation for the delay,
\begin{equation}\label{eq:tauST}
    \exp{\left({-\sqrt{3} \text{Oh}_c}\tau_{ST}\right)} \sinh\left(\sqrt{3}\text{Oh}_c\sqrt{\frac{1}{2} + \text{Oh}_c^2}\tau_{ST} \right) = \frac{\sqrt{\frac{1}{2} + \text{Oh}_c^2}}{5\sqrt{3}\varepsilon}.
\end{equation}
In the limit case of undamped growth, one may extract an explicit term 
\begin{equation}\label{eq:corellUN}
    \tau_{ST} \approx \sqrt{\frac{2}{3}} \sinh^{-1}\left({\frac{1}{5\sqrt{6}\varepsilon}}\right).
\end{equation}
Using the form of \cref{eq:corellUN} and numerical solution of \cref{eq:tauST}, we derived an explicit correlation for the surface tension delay under generalized, damped growth:
\begin{equation}\label{eq:corellDAM}
\tau_{ST} \approx \left( {1+ 2.849 \textcolor{black}{\text{Oh}_{c}} \varepsilon^{0.001}} \right)\sqrt{\frac{2}{3}} \sinh^{-1}\left({ \frac{\sqrt{\frac{1}{2}+\text{Oh}_c^2}} {5\sqrt{3}\varepsilon} } \right).
\end{equation}

\begin{figure}
    \centering 
    \includegraphics[width=.6\columnwidth]{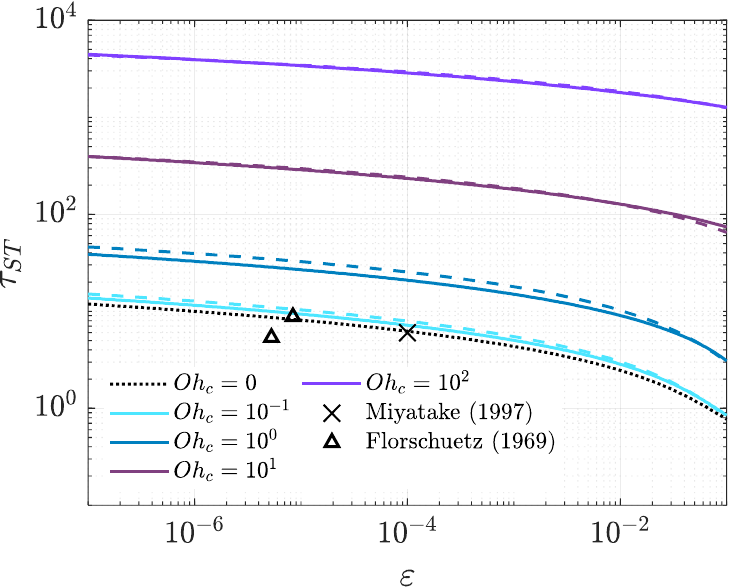}
    \caption{Surface-tension-induced delay as a function of the initial perturbation $\varepsilon$ for various values of nucleation Ohnesorge number $\text{Oh}_{c}$. Solid lines demarcated the delay extracted using the third-order perturbation solution, whereas dashed lines correspond to the prediction of the correlation \cref{eq:corellDAM}. A black dotted line demarcated the limiting case of undamped growth $\text{Oh}_c=0$. The surface tension delay reported by \citeauthor{miyatake1997}\citep{miyatake1997} and extracted from the experimental results of \citeauthor{florschuetz1969}\citep{florschuetz1969} is indicated by a black cross and triangles, respectively.} 
    \label{fig:DampedTauEps}
\end{figure}
The surface-tension-induced delay is set by both the nucleation Ohnesorge number and the initial perturbation from the stable radius, with a shorter delay expected for larger deviations from equilibrium. 
The relation between the two is illustrated by \cref{fig:DampedTauEps}, presenting the surface tension delay as a function of the normalized perturbation $\varepsilon$ for various values of nucleation Ohnesorge number \textcolor{black}{$\text{Oh}_{c}$}.
\Cref{fig:DampedTauEps} reveals that delay values for low \textcolor{black}{Ohnesorge} values ($\text{Oh}_c = 0$--$10^{-1}$) converge as expected, indicating that the delay is primarily controlled by surface-tension forces dictated by the initial perturbation.
In this range, the delay exhibits higher relative dependence on $\varepsilon$ -- an order of magnitude change. 
The transition in dominant delaying mechanisms around $\text{Oh}_c = 10^{-1}$--$10^0$ also alters the sensitivity to the initial perturbation, with a small relative change at higher Ohnesorge values ($\text{Oh}_c = 10^{1}$--$10^{2}$). 
Yet significant absolute changes in the delay are still expected at these values, highlighting that the perturbation magnitude could not be overlooked when viscosity is the dominant delaying mechanism.
The extended correlation, presented here in \cref{eq:corellDAM}, proves accurate for high Ohnesorge values but overestimates the delay, especially for small initial perturbations.
Although the definition of the delay is not deterministic and aims to evaluate only the phenomenon's order of magnitude, the correlation could still provide a prompt estimation of the delay. 
While limited in its extent, comparing the available data on initial delay reveals its agreement with our model's predictions.
\citeauthor{miyatake1997}~\citep{miyatake1997} reported, while normalized in terms of our analysis, a delay of $\tau_{ST}=6$ for an initial perturbation of $\varepsilon=10^{-4}$.
This result was obtained by correlating the undamped analytical, numerical, and experimental results. 
As illustrated in \cref{fig:DampedTauEps}, our findings corroborate this observation: this singular value and the predictions derived from our model for $\text{Oh}_c=0$ agree well.
Surface tension delay values were also extracted from the raw data reported by \citeauthor{florschuetz1969}~\citep{florschuetz1969}.
While our model and these results exhibit some discrepancies, it suggests that the framework developed herein offers a viable approximation for the delay timescale.

\subsection{Unified solution}\label{sec:unif}
While the inner solution validity is limited to the near-equilibrium bubble radii, its matching outer solution may expand the analysis region of validity to larger radii. 
The unified solution elucidates three distinct stages in the bubble's growth and provides a new analytical approximation for the surface-tension-dominated and transitional regimes.
\begin{figure*}
    \centering 
    \includegraphics[width=\linewidth]{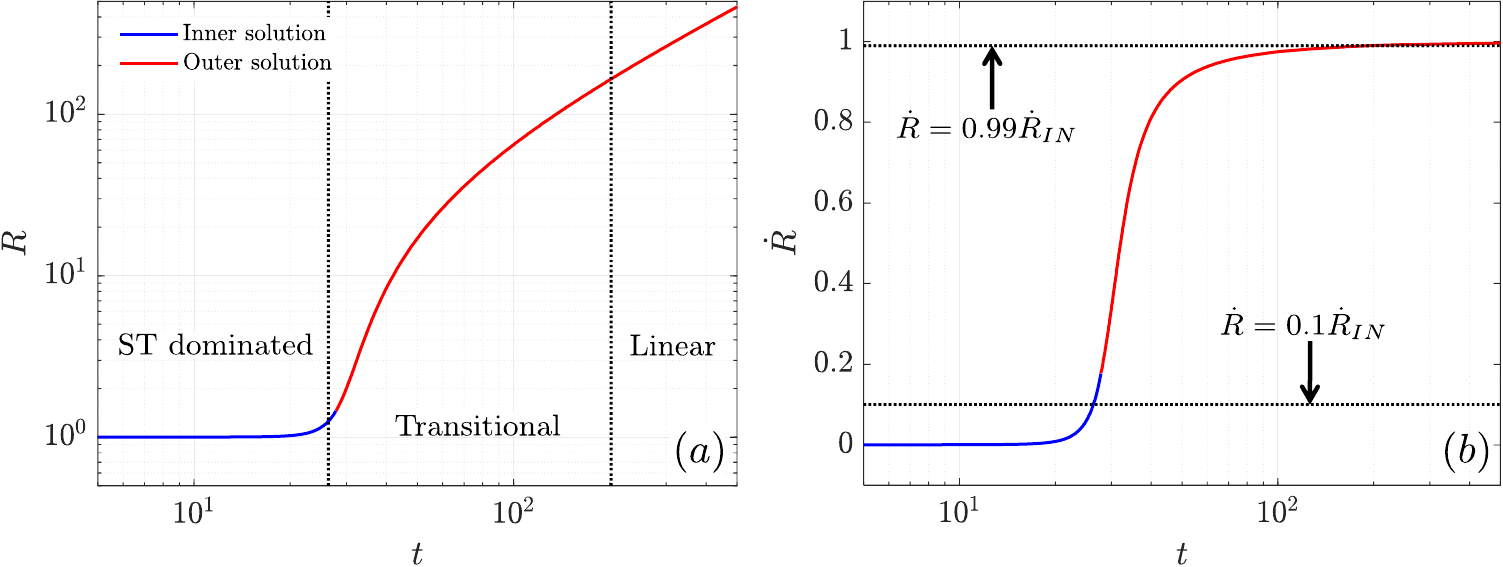}
    \caption{Unified approximated solution illustrating (a) bubble radius as a function of time and (b) bubble velocity as a function of time of a vapor bubble. The initial perturbation is $\varepsilon = 10^{-5}$ and the nucleation Ohnesorge number is $\text{Oh}_c=1$. The inner, first-order solution, given by \cref{eq:sol1}, is depicted by a blue line. The matching outer solution, given by \cref{eq:outer}, is depicted by a red line. Dotted vertical lines demarcate the growth regimes: a surface-tension-dominated regime, defined using velocity threshold $\dot{R}<0.1$, a linear regime, defined using velocity threshold $\dot{R}>0.99$, and an intermediary, transitional regime.}
    \label{fig:UnAnalyticSol}
\end{figure*}
The unified solution, \cref{eq:sol1} (inner solution) and \cref{eq:outer} (outer solution) prediction for the test case ($\text{Oh}_c=1$,$\varepsilon = 10^{-5}$) is presented in \cref{fig:UnAnalyticSol}.
As previously discussed, the inner solution predicts delayed growth due to the dominance of surface tension forces, starting its transition to the inertial, linear growth regime as the velocity threshold $\dot{R}\leq0.1$ is crossed. 
In the transitional regime, the bubble velocity increases sharply, as it tends to the inertial velocity limit $\dot{R} \rightarrow 1$.
The end of the transitional regime is marked by a second threshold, 
after which the bubble's velocity is practically constant, as expected for inertially limited growth. 
The criterion to distinguish between the two is arbitrarily set here to $\dot{R}=0.99$, demarcating the transition into the fully inertial regime.
We offer this criterion to provide a quantifiable indicator, facilitating a distinct separation between the growth regimes.
\begin{figure}
    \centering 
    \includegraphics[width=.6\columnwidth]{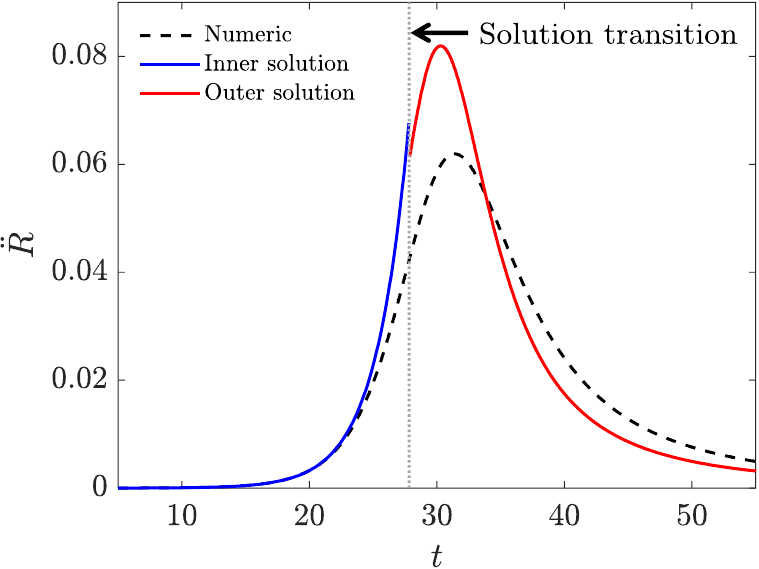}
    \caption{Bubble acceleration as a function of time: a comparison between numerical and analytically obtained approximations for initial perturbation of $\varepsilon = 10^{-5}$ and nucleation Ohnesorge number of $\text{Oh}_c=1$. The solution transition is indicated by a dotted vertical line.}
    \label{fig:RddotUnif}
\end{figure}

\Cref{fig:RddotUnif} depicts the bubble acceleration as a function of time, presenting a comparison between the unified solution approximation and the numerical solution.
Notably, the trade-off inherent to simplifying the original RP oscillator is revealed by \cref{fig:RddotUnif}.
A discontinuity in acceleration arises when transitioning from the inner to the outer solution, illustrated by a dotted vertical line.
The unified solution also overestimates the acceleration during peak acceleration and undershoots at long timescales.
Despite these deviations, the pattern of the unified solution mirrors that of the numerical solution, validating its utility as a reasonable approximation, especially for the surface-tension-dominated regime.

\begin{figure*}
    \centering 
    \includegraphics[width=\linewidth]{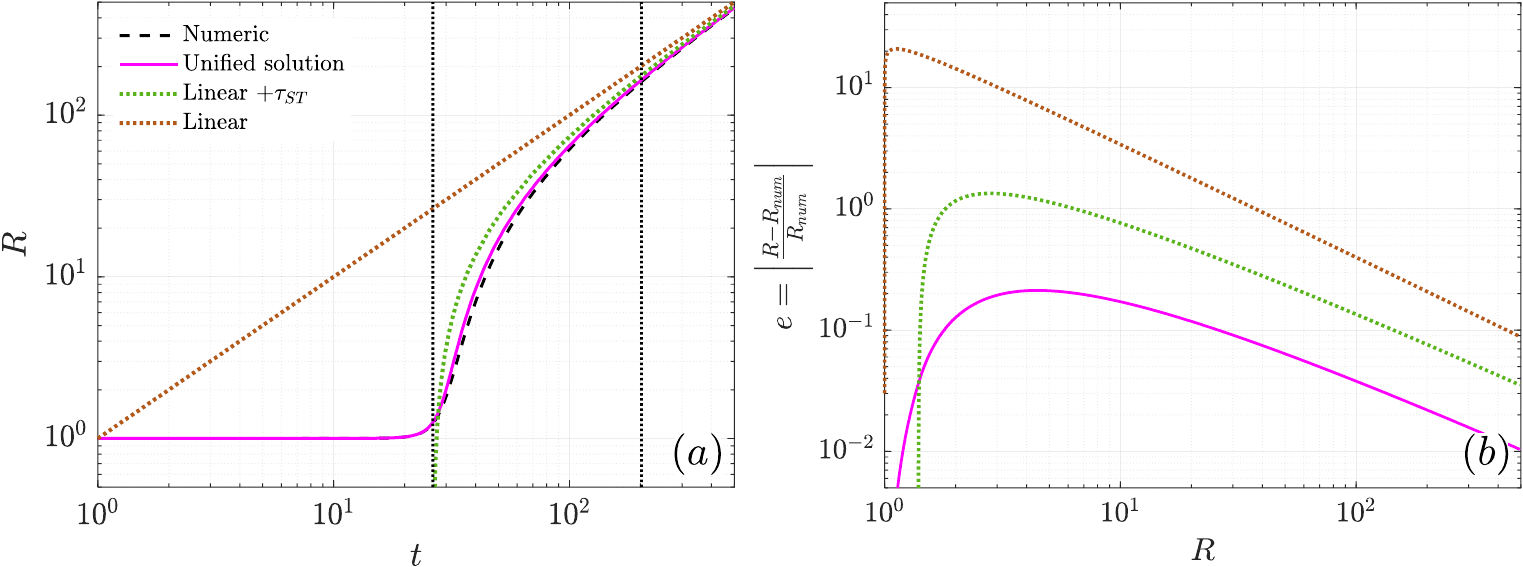}
    \caption{A comparison of growth models (a), and their associated relative error with respect to the numerical model $e = \lvert  {\frac{R-R_{num}}{R_{num}}} \rvert$ (b).
    The dashed black line depicts the numerical solution, while the pink line represents the unified model, both derived for an initial perturbation of $\varepsilon = 10^{-5}$ and nucleation Ohnesorge number of $\text{Oh}_c=1$.
    The dotted brown line illustrates the estimation of conventional models, such as the MRG model \citep{Mikic1970}, in which the surface tension delay is neglected, and the bubble is assumed to begin its growth in the inertial regime.
    The green dotted line incorporates the surface tension delay, retarding the linear growth to account for the initial delay.}
    \label{fig:COMP}
\end{figure*}
\Cref{fig:COMP} provides a comparative analysis of different growth models. 
Standard analytical models such as the MRG model often overlook the initial surface-tension-induced delay and assume the growth begins from the linear regime -- as depicted by the dashed brown line.
We propose two enhancements for the standard analytical models: the first is the unified solution (solid pink line), and the second involves incorporating the surface tension delay time to retard the onset of linear growth (dashed green line).
This simple retardation immediately improves the model's agreement with the numerical result.
However, as illustrated in \cref{fig:COMP}, the uniform solution offers a markedly more accurate approximation for short time scales. 
The deviation between the uniform solution and the numerical result stems from excluding the unsteady inertia term, notably impacting the intermediate regime.
The errors relative to the numerical model are depicted in \cref{fig:COMP}(b), revealing that the linear model significantly overestimates the bubble size within timescales of up to 100.
Introducing a simple retardation immediately enhances accuracy by an order of magnitude. 
Notably, all models converge at longer timescales, exhibiting a linear decrease in their relative error.

Overall, the findings demonstrated in \cref{fig:COMP} emphasize the significance of considering the surface tension delay, particularly during short timescales, and highlight its crucial role in accurately modeling the dynamic behavior of the bubble. 
In normalized terms and assuming the relation between the vapor pressure and temperature can be expressed by the linearized Clausius-Clapeyron equation, the MRG model \citep{Mikic1970} predicted that the timescale in which thermal effects begin to affect the bubble growth is
\begin{equation}
    \tau_{TH} = \frac{18\alpha_l}{\pi\sigma} \left(\frac{c_{p,l} T_{\text{sat}}}{\rho_v^2 h_{\text{fg}}^2}\right)^2 (\rho_l \Delta p_0)^{5/2},
\end{equation}
where $\alpha_l$,$c_{p,l}$ are the liquid heat diffusivity and capacity, correspondingly, $T_{sat}$ is the saturation temperature at the liquid far-field pressure, $h_{fg}$ is the latent heat of vaporization and $\rho_v$ is the vapor density.
Comparing $\tau_{TH}$ magnitude with $\tau_{ST}$ magnitude illuminates the thermal effects timescale and could suggest whether the isothermal assumption made in our model is justified. 
If $\tau_{ST}$ is approximately equal to $\tau_{TH}$, then thermal effects modify the bubble's behavior even during the initial delay.
\Cref{fig:COMP} indicates that the transition to fully inertial growth occurs after the growth reaches approximately $\tau_{IN}\approx10\tau_{ST}$, implying that fully inertial growth can only be sustained when the thermal timescale significantly exceeds the timescale associated with surface tension effects $\tau_{TH}>>\tau_{ST}$.
Otherwise, the bubble will not stabilize at a constant growth velocity as a direct transition from surface tension regime to thermal regime may occur.
$\tau_{TH}$ depends on the superheated liquid's thermochemical properties and the conditions at the onset of the phase change, serving as a simple yet powerful parameter allowing one to assess the main mechanisms dictating the early bubble growth stages.
For example, high liquid superheats result in a relatively long inertial growth regime, while low surface tension shortens the initial delay, thus changing the ratio between the thermal timescales and the other physical mechanisms governing the growth.
High heat capacity and thermal diffusivity tend to retard the vapor's thermal response, whereas a high latent heat accelerates the vapor's temperature drop to $T_{\text{sat}}$.
This contrasting effect highlights the complex interplay between the underlying physical mechanisms influencing bubble dynamics and growth behavior.

\section{Conclusions}\label{sec:conc}
This study focuses on a near-equilibrium, surface-tension-dominated growth regime of vapor bubbles and its transition to an inertia-dominated regime.
We adopted a regular perturbation method to analytically derive an approximate inner solution that was then matched with an outer solution for the transitional regime.
The unified solution offers a closed mathematical formulation for the first time for the early growth stages of vapor bubbles.
The analytical solutions were validated with direct numerical solutions of the Rayleigh-Plesset equation, successfully capturing the initial growth delay.
Our analysis predicts the initial delay period as a function of two nondimensional parameters: the normalized perturbation from criticality and the nucleation Ohnesorge number, encapsulating the interplay between viscous damping and surface tension forces near criticality. 
We found that viscous effects play a significant role in the near-equilibrium vapor bubble growth: it may dampen the bubble's expansion, extending the surface-tension-dominated regime and retarding the transition to the inertial regime.
For low nucleation Ohnesorge numbers, $\text{Oh}_c<1$, the delay is governed by surface tension, while higher values, $\text{Oh}_c>1$, lead to a linear correlation with viscous damping.
The comparative analysis with standard growth models emphasizes the importance of considering surface tension delay, particularly at short timescales, for accurate modeling of bubble behavior.

The developed unified solution suggests a quantifiable indicator facilitating a distinct separation between the three growth stages.
The derived analytical solutions, particularly the correlation for surface-tension-induced delay, may be integrated into existing models for vapor bubble growth.
Furthermore, a simple parameter, $\tau_{ST}$, was derived to estimate the extent to which thermal effects alter the bubble dynamics during the early growth stages.
Altogether, the results presented here offer a rigorous yet simple method to dissect the complex interplay between surface tension, viscosity, inertia, and thermal effects, suggesting that understanding the relative timescales of these mechanisms is crucial for predicting bubble behavior accurately.

The obtained results highlight the necessity of further experimental efforts to study post-nucleation vapor bubble dynamics to capture the predicted growth delay. 
Finally, this approach may be extended to include bubble curvature-dependent and dynamic surface tension models, unsteady external pressure excitation, and gas content effects. 
Such an extension broadens the applicability and enhances model predictions, especially applicable to bubbles nucleating within highly superheated liquids; an ongoing study aims at unraveling the influence of these phenomena on near-equilibrium dynamics.

\begin{acknowledgments}
The authors are indebted to Dr. Tali Bar-Kohany for her critical advice and fruitful discussions. This research was supported by the ISRAEL SCIENCE FOUNDATION (grant No. 1762/20). ES acknowledges the financial support of Minerva Research Center (Max Plank Society Contract No. AZ5746940764). 
\end{acknowledgments}

\nomenclature{${R}$}{Bubble radius}
\nomenclature{${R}_c = \frac{2\sigma}{\Delta p_0}$}{Critical bubble radius}
\nomenclature{$\sigma$}{Vapor-liquid surface tension}
\nomenclature{$\Delta p$}{Pressure gradient across the bubble}
\nomenclature{$\varepsilon$}{Initial radius perturbation from critical bubble radius}
\nomenclature{$n_e$}{Equilibrium embryo distribution function}
\nomenclature{$k$}{Boltzmann's constant}
\nomenclature{$T$}{Temperature}
\nomenclature{$\mu_l$}{Liquid viscosity}
\nomenclature{$\rho_l$}{Liquid density}
\nomenclature{$\text{Oh}_c$}{Nucleation Ohnesorge number}
\nomenclature{$R_0, R_1, R_2, R_3$}{Perturbation series terms}
\nomenclature{$t$}{Time}
\nomenclature{$C_1$}{Integration constant}
\nomenclature{$\tau_{ST}$}{Surface tension delay time}
\nomenclature{$\tau_{TH}$}{Timescale associated with thermal effects}
\nomenclature{$\tau_{IN}$}{Timescale associated with inertial effects}
\nomenclature[O]{$\tilde{}$}{Dimensional variable}
\nomenclature[O]{$\dot{}$}{First order time derivative}
\nomenclature[O]{$\ddot{}$}{Second order time derivative}
\nomenclature[S]{in}{Inner solution}
\nomenclature[S]{out}{Outer solution}
\nomenclature[S]{tan}{Value at the tangent
intersection of the inner and outer solutions}
\nomenclature[S]{num}{Numerical solution}
\nomenclature{$\alpha_l$}{Liquid heat diffusivity}
\nomenclature{$c_{p,l}$}{Liquid heat capacity}
\nomenclature[S]{sat}{Value at saturation}
\nomenclature{$h_{\text{fg}}$}{Latent heat of vaporization}
\nomenclature{$\rho_v$}{Vapor density}


\bibliography{Main.bib}

\end{document}